\documentclass[pdflatex, sn-vancouver-num]{sn-jnl}

\usepackage{graphicx}%
\usepackage{multirow}%
\usepackage{amsmath,amssymb,amsfonts}%
\usepackage{amsthm}%
\usepackage{mathrsfs}%
\usepackage[title]{appendix}%
\usepackage{xcolor}%
\usepackage{textcomp}%
\usepackage{manyfoot}%
\usepackage{booktabs}%
\usepackage{algorithm}%
\usepackage{algorithmicx}%
\usepackage{algpseudocode}%
\usepackage{listings}%
\usepackage{natbib}
\usepackage{comment}
\usepackage{xurl}

\usepackage{cleveref}

\begin{document}

\title[From Chat Control to Robot Control]{From Chat Control to Robot Control:\\  Implications of the Chat Control Proposal for Human-Robot Interaction}
\author*[1]{\fnm{Neziha} \sur{Akalin}}\email{neziha.akalin@ju.se}
\equalcont{These authors contributed equally to this work.}

\author[2]{\fnm{Alberto} \sur{Giaretta}}\email{alberto.giaretta@oru.se}
\equalcont{These authors contributed equally to this work.}

\affil*[1]{\orgdiv{Department of Computer Science and Informatics}, \orgname{J\"onk\"oping University}, \orgaddress{\street{Gjuterigatan 5}, \city{J\"onk\"oping}, \postcode{551 11}, \country{Sweden}}}

\affil[2]{\orgdiv{Department of Computer Science}, \orgname{Örebro University}, \\\orgaddress{\street{Fakultetsgatan 1}, \city{Örebro}, \postcode{70182}, \country{Sweden}}}


\abstract{This paper explores how a recent European Union proposal, the so-called Chat Control, which creates regulatory incentives for providers to implement content detection and communication scanning, could transform the foundations of human–robot interaction (HRI). As robots increasingly act as interpersonal communication channels in care, education, and telepresence, they convey not only speech but also gesture, emotion, and contextual cues. We argue that extending digital surveillance laws to such embodied systems would entail continuous monitoring, embedding observation into the very design of everyday robots. This regulation blurs the line between protection and control, turning companions into potential informants. At the same time, monitoring mechanisms that undermine end-to-end encryption function as de facto backdoors, expanding the attack surface and allowing adversaries to exploit legally induced monitoring infrastructures. This creates a paradox of safety through insecurity: systems introduced to protect users may instead compromise their privacy, autonomy, and trust. 
This work does not aim to predict the future, but to raise awareness and help prevent certain futures from materialising.}


\newcommand{\qt}[1]{``#1''}

\maketitle

\section{Introduction}\label{sec1}

In 2020, the European Union (EU) and Europol reported a sharp increase in online child sexual abuse material (CSAM)~\cite{europol2020}. Spurred by the urgency of the matter, on 11 May 2022 the EU Commission proposed a new regulation (Proposal 52022PC0209~\cite{EURLexce65:online}) on the prevention and combat of child sexual abuse~\cite{Commissi67:online}. 
In its original form, the original Commission proposal granted authorities the power to compel providers of Interpersonal Communication Services (ICS) including apps and social media platforms (such as WhatsApp and Instagram) to scan all communications, including encrypted ones, across text, image, and video. Within the original Commission proposal, the same framework applied to hosting service providers, such as Dropbox, and continues to apply in substance in the Council of the European Union's revised position, as discussed in this paper. Cybersecurity researchers have characterised such detection orders and client-side scanning as neither effective nor surveillance-preventing~\cite{abelson2024}, and as \textit{de facto} backdoors~\cite{shurson2024}. Over 800 European experts signed a joint statement raising serious concerns about the proposal~\cite{jointauthors2025}. Expressing similar concerns, digital rights groups~\cite{gff,edri2025} and Members of the European Parliament~\cite{fragkos2025} have labelled the initiative mass surveillance incompatible with fundamental rights. Stemming from these observations, the proposal has been nicknamed Chat Control.

Under the Danish Presidency, in October 2025 the Council consolidated Member State positions on the proposal, with a majority expressing support. A Council vote was scheduled for mid-October 2025 \cite{eucrypto2025}, but was postponed amid growing concerns. As the legislative process progressed, the proposal underwent further revision. In December 2025, the Council adopted a revised position that removed the explicit mandate and reframed such measures as formally voluntary, while retaining the regulatory framework based on risk assessment, mitigation duties, and provider responsibility~\cite{consiliumeuropa2025}. The proposal has now entered inter-institutional negotiations (trilogue) between the European Parliament, the Council, and the Commission, with no final legal text yet adopted~\cite{europalegfight2026}. Despite the removal of an explicit mandate, we argue that the revised framework continues to incentivise broad monitoring practices in order to demonstrate compliance and limit regulatory exposure. 


The serious interference with EU citizens' fundamental rights has been extensively analysed since the proposal was rendered public. In this paper, we examine how the proposal's broad definition of ICS could potentially extend to systems that facilitate human communication, in particular socially assistive robots. 
These robots are increasingly used, in various research initiatives and commercial deployments, to engage in natural language conversations with users across multiple applications. Notable examples are nursing care \cite{nieto2022survey}, hospitals, elderly care centers, occupational health facilities, private homes, and educational institutions \cite{aymerich2023socially}. 
In the near future, providers of human-robot interaction (HRI) solutions might be pressured to equip their robots with monitoring backdoors to comply with Chat Control. As robots are typically equipped with multiple sensors, including cameras and microphones, their deployment in contexts such as remote patient assistance and monitoring could lead service providers to implement extensive monitoring practices to meet regulatory obligations. This, in turn, may result in the collection and analysis of highly personal and intimate aspects of individuals' lives.

Potential users of socially assistive robots have already expressed concerns regarding privacy intrusion through monitoring, data ethics, human agency, perceived loss of control, and the management of personal data~\cite{fronemann2022should}. Researchers have consistently emphasized the importance of addressing these concerns to support the responsible integration and acceptance of robots in everyday life \cite{nieto2025robot}. Regulatory measures that mandate extensive monitoring practices, such as those proposed under the Chat Control proposal, risk exacerbating these existing concerns rather than alleviating them.

\section{The Chat Control Proposal}
The proposed CSAM regulation establishes a set of risk-based compliance mechanisms applicable to ICS providers. While multiple provisions of the original Commission proposal raised concerns, we focus on the regulatory logic articulated most explicitly through Articles 3, 7, and 10. Although Articles 7 and 10 were removed in the Council’s revised position, their analysis remains crucial for understanding the nature of the current risk-based framework.

Article 3 establishes risk assessment measures, stating that \qt{providers of interpersonal communication services shall identify, analyse and assess [..] the risk of use of the service for the purpose of online child sexual abuse}. More in detail, Article 3(2)(e).(i) states that providers shall take into account, in particular \qt{the extent to which the service is used or is likely to be used by children}, when carrying out a risk assessment. In addition, Article 3(2)(e).(iii) states that providers shall take into account to which extent their service enables \qt{users to establish contact with other users directly, in particular through private communications}. In principle, these are all commendable goals. 

However, Article 3(5) states that such risk assessment shall \qt{include an assessment of any potential remaining risk that, after taking the mitigation measures pursuant to Article 4, the service is used for the purpose of online child sexual abuse}. Unfortunately, the complete elimination of residual risk is technically impossible. Even the most advanced machine-learning models cannot achieve perfect accuracy or completely avoid false positives and false negatives. While detection systems continue to improve, no viable solution can guarantee that all CSAM content will be identified without error.

As a consequence, this provision establishes a \textit{de facto} perpetual state of liability for ICS providers. Article 3(6) compounds the problem, stating that \qt{The Commission, in cooperation with Coordinating Authorities [..], may issue guidelines on the application of paragraphs (1) to (5), having due regard in particular to relevant technological developments [..]}. Combined, paragraphs (5) and (6) risk creating a never-ending loop of compliance burden for ICS providers. Since the risk that CSAM material may bypass detection can never be eliminated, each new round of mitigation could justify demands for stricter interventions, perpetually driven by the persistence of residual uncertainty. Proposal advocates might argue that the goal is merely to keep improving detection systems. But that view depends on the assumption that it will be enforced fairly and that the limits of its use will always be clear. The proposal risks transforming risk assessment from a bounded technical process into an open-ended compliance obligation, with significant implications for providers that design and deploy ICS.

In addition, the Council's revised position has introduced a new Article 3(4a), which states that \qt{the risk assessment shall gather information on the limitation of the risk to an identifiable part or component of the service where possible [..] or to specific users or specific groups or types of users where possible, to the extent that such part, component, specific users or specific groups or types of users can be assessed in isolation [..]}. Although this provision is framed as a measure that limits the extent of scrutiny, it functions in practice as a compliance burden. By suggesting that risks can be bounded, Article 3(4a) imposes on providers the obligation to attempt the localisation, segmentation, and classification of such risks. In other words, Article 3(4a) does not operate as a narrowing mechanism, but rather as an additional infrastructural obligation.

Although Articles 7 and 10 were removed in the Council's revised position, it remains critical to analyse them as an explicit articulation of the compliance logic that complements and clarifies the risk-based framework centred on Article 3. Article 7(1) stated \qt{The Coordinating Authority of establishment shall have the power to request the competent judicial authority of the Member State [..] to issue a detection order requiring a provider [..] to take the measures specified in Article 10 to detect online child sexual abuse on a specific service.} This provision entailed that judicial authorities would have been granted the power to enforce the use of such measures. 

Concerning the specific measures, Article 10(1) read \qt{Providers [..] that have received a detection order shall execute it by installing and operating technologies to detect the dissemination of known or new child sexual abuse material or the solicitation of children}. However, selective detection without broad-scale data collection and analysis is technically infeasible. To determine whether any given content may constitute CSAM (or any other category, for that matter), the entire corpus of user communications must first be collected and examined. Consequently, to comply with Article 10, providers would have been effectively forced to collect all users' communications. In other words, mass surveillance would not have been an unintended side effect of the proposal, but its \textit{sine qua non} condition.

In this section, we have covered some troublesome passages within the relevant articles, as they stood in the Commission proposal and as they remain in the Council's revised position. In the next section, we discuss how the formulation, intended to regulate communication platforms, could be leveraged to influence HRI services.

\section{From Street Cameras to Living Rooms}
Modern societies have already normalized extensive public surveillance. Cameras placed on streets, in public transport, and across urban infrastructures are routinely justified as tools for safety, crime prevention, and efficiency~\cite{piza2019}, although the debate on their effectiveness is open~\cite{welsh2002,nieto2002}. Citizens have grown accustomed to being watched in public, often accepting visibility as the price of security. Over time, these technologies have expanded not only in coverage but in capability. Researchers speculate that, by combining facial recognition with behavioural prediction, future CCTV systems could evolve toward proactivity, enabling forms of automated law enforcement~\cite{skogan2019}. 

\begin{figure}[t]
  \centering
  \includegraphics[width=0.8\linewidth]{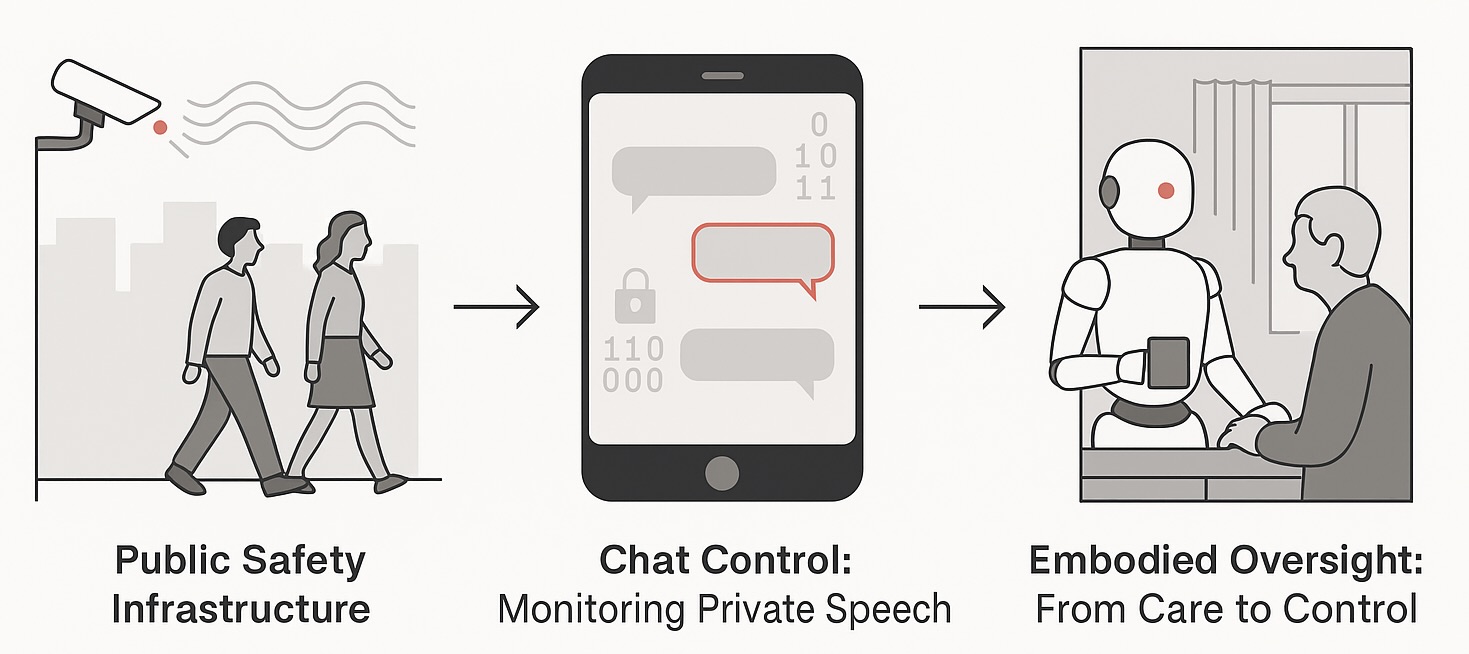}
  \caption{The continuum of surveillance, from watching public spaces, to listening in private communications, to acting within embodied environments.}
  \label{fig:continuum}
\end{figure}

Both indoor and outdoor surveillance robots have been developed with the explicit goal of providing security monitoring services~\cite{chun2016robot}. In contrast, care robots with monitoring capabilities, especially when deployed in intimate spaces, have been subject to ethical debate since the early 2010s~\cite{sharkey2012granny}. The Chat Control proposal may further extend this trajectory by shifting the focus of surveillance, from predominantly public spaces toward private environments, as illustrated in~\Cref{fig:continuum}. Such a development would constitute not only an expansion in the scale of surveillance practices, but also a transformation in their nature, extending from the monitoring of public activities to the scrutiny of routine interactions within private spaces.


As discussed in the following section, the Chat Control proposal is deliberately framed in broad and flexible terms. This breadth makes it applicable to a significantly wider set of scenarios than its proponents explicitly anticipate or admit.

\section{From Messaging Apps to Embodied Systems}
As we have previously discussed, the proposed legislation is framed in such a way that authorities could, in principle, be driven to escalate intrusive monitoring measures. On top of that, the proposal has a critical weakness in its attempt to regulate ICS in general. 

The issue at hand is that the proposal imports the definition of ICS from the European Electronic Communications Code (EECC), Directive (EU) 2018/1972. In particular, Article 2(5) EECC defines ICS as \qt{a service [..] that enables direct interpersonal and interactive exchange of information via electronic communications networks [..]}. ICS constitutes a deliberately broad legal category, encompassing any service that enables users to exchange messages over a network. The definition continues with \qt{[ICS] does not include services which enable interpersonal and interactive communication merely as a minor ancillary feature that is intrinsically linked to another service}. Although this clause excludes services where communication is limited and incidental, various robots explicitly designed to mediate interaction fall under the ICS category; socially assistive robots used in elder care (audio/video contact), telepresence robots in healthcare, or service robots enabling communication in schools could all qualify as ICS. 

For example, Celsius School in Edsbyn, Sweden, in 2023 started offering robot Otto to students who are unable to show up for school~\cite{svtOtto}. Otto (technical name, AV1) is a robot produced by the Norwegian start-up company No Isolation, created to facilitate learning and social contact by acting as an embodiment of the child in the classroom~\cite{noisolation}. In December 2021, more than 1600 units were active across Europe~\cite{johannessen2023student}. The robot allows students to participate remotely in classrooms and communicate with the teacher through the robot's facial expressions. For the child using it, Otto is not a mere communication medium. It is their temporary embodiment, representing their own existence within the collegial classroom experience. Once a legal mechanism proposed to protect the pupil's safety, Chat Control becomes the tool that dissects their words and thoughts.

It is worth noting that, although the proposal imports the ICS definition from the EECC, its Article 2(b) adds a clause on the applicability of the regulations on \textit{publicly available} service. The term \textit{publicly available} is not formally defined in the EECC, but its Recital 16 (as well as subsequent regulatory interpretations) clarify that it refers to services offered to end-users in general, rather than to a closed or exclusive group. Under this definition, any commercial robot available on the free market and purchasable by any customer under the same general terms constitutes a publicly available service. As an example, the Double 3 robot from Double Robotics is a self-driving, video-conferencing robot designed for working remotely in hybrid offices and classrooms~\cite{double3}. Although it requires purchase and service subscription, this product is available to the public and therefore qualifies as a publicly available service.

Beyond interpersonal communications we covered so far, the proposal seeks to regulate another category of service providers, namely hosting service providers. Chat Control relies on the Digital Services Act (DSA) to define hosting service. In particular, Article 3(g) of the DSA states that hosting consists of \qt{the storage of information provided by, and at the request of, a recipient of the service}. 

When users interact with robots, there are cases in which the interaction could be configured as request storage, albeit implicit. For example, if a healthcare telepresence robot allows the user to record a video note for their caregiver hosted on the Cloud, the user's action triggers a \textit{de facto} request. Even if a robotic platform does not fall under the definition of ICS, it could still be regulated by Chat Control if it just stores any kind of information connected to HRI.

It is true that the clause \qt{at the request of} narrows the scope of the provision. Some may argue that this clause excludes service robots that log events (and store them on the Cloud) for performing diagnostics, as automated logging cannot be considered triggered by a user's request. However, when it comes to robots (or any other embodied system), the distinction between automated logs and user-provided content is blurry, by definition. The act of engaging with an embodiment shows the voluntary intent of interacting with it. The consequent automatic logging may be argued to arise from user-initiated interaction flows in some designs, hence configure as an implicit request.


In conclusion, the regulation does not appear to have been drafted with embodied systems in mind, as neither robots nor embodied systems are explicitly addressed in the proposal or its explanatory memorandum. However, given its intentionally broad and future-oriented scope, the regulation could potentially encompass such systems as they evolve and become further integrated into everyday life. This raises concerns that the proposed measures may enable forms of monitoring that extend into increasingly private domains of interaction. Content regulation that is presented as a means of protecting vulnerable groups may, in practice, lead to expanded oversight of intimate communicative practices, thereby reshaping the boundaries between public regulation and private life.

\section{Turning Robots from Companions to Censors}
In the previous sections, we showed that the regulatory language of Chat Control could already encompass embodied systems such as service and social robots. 
We now analyse how a legal framework originally designed for disembodied digital communication may operate when applied to systems that are physically situated, sensor-equipped, and interactive. Social and humanoid robots constitute a distinct form of interpersonal communication medium. They do not simply transmit information; rather, they mediate presence, emotion, and intention, shaping how individuals communicate across distance~\cite{zhao2006humanoid}. In doing so, they transform communication into a form of embodied co-presence that extends beyond the mere exchange of data. Interpreting such systems solely as data transmission channels, as the Chat Control proposal may implicitly suggest, overlooks this relational dimension and risks turning this companionship into compliance.

Socially assistive robots have been proved to be effective in hospitals, elderly care centers, occupational health centers, private homes, and educational institutions \cite{aymerich2023socially}. Telepresence robots enable remote participation in classrooms~\cite{johannessen2023student}, workplaces (e.g. Double robot \cite{double3}), and healthcare settings, thereby allowing users to project aspects of their physical presence into distant environments. Likewise, remote-controlled or semi-autonomous robots mediate interaction across distance through embodied engagement. From a regulatory standpoint, this communicative capacity may place social and telepresence robots within the potential scope of the Chat Control framework. As these robots enable the exchange of information between individuals, the dialogue facilitated by them, hence the care and companionship components of such exchange, could become subject to similar surveillance or content-scanning obligations proposed for traditional messaging platforms. Such obligations would reshape the role of these systems, shifting them from trusted interaction partners to intermediaries that are legally required to monitor and potentially report user communications. 

Although artificial intelligence (AI) has so far remained implicit in this discussion, AI-enabled robots further complicate the regulatory landscape. In addition to embodying communication, these systems interpret, generate, and adapt communicative content autonomously. They may infer user intentions, estimate affective states, and personalize responses. When such systems are subject to detection or reporting requirements, monitoring may extend beyond passive data transmission to include the automated interpretation of meaning and context. 

\begin{figure}[t]
  \centering
  \includegraphics[width=0.8\linewidth]{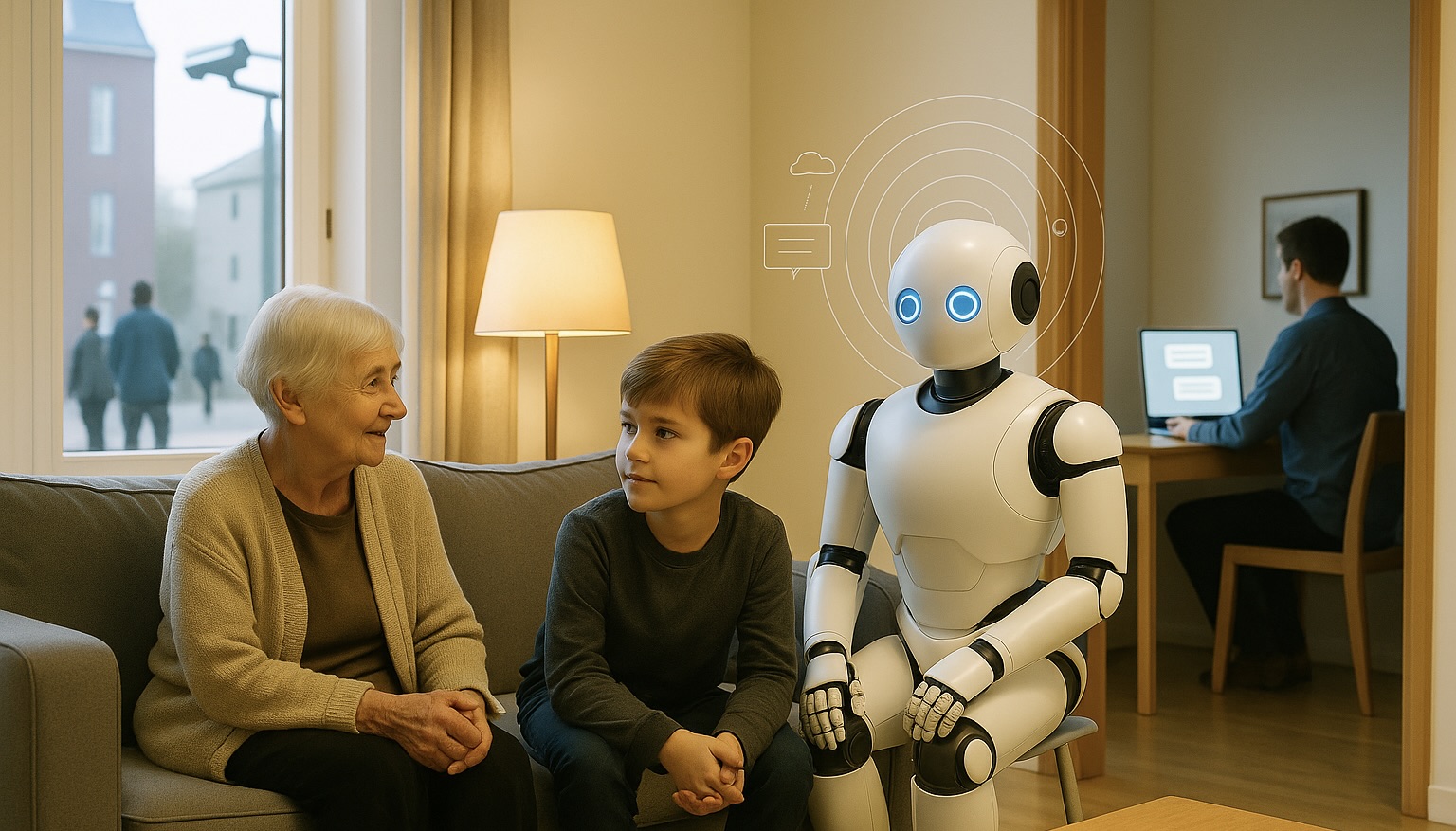}
  \caption{Conceptual illustration of a speculative future where domestic robots, designed for care and companionship, become instruments of observation. Image generated by the author using ChatGPT (GPT-5) image generation tools.}
  \label{fig:watching}
\end{figure}

A care robot that provides comfort to an elderly person might also record their expressions to detect signs of \qt{risk}. Telepresence robots designed to connect family members and marketed as privacy-sensitive could, under a legal framework, audit conversational logs. In each case, the same interaction that enables empathy and trust simultaneously produces a data trace that may later serve as evidence, training material, or security telemetry. If this logic remains unchecked, the future of HRI may resemble a form of ambient regulation, where laws are not only enforced through institutions but embedded directly into the technical architectures of communication. Robots would not simply comply with regulatory frameworks; they would enact them, becoming both the medium and the control mechanism. The danger is not that such systems would immediately resemble dystopian enforcers, but that they would quietly normalize invisible observers within intimate spaces. In other words, that they would become backdoors themselves. This dynamic is conceptually illustrated in~\Cref{fig:watching}, which envisions domestic robots silently observing an elderly person and a child in a shared space.

To prevent this, policymakers and designers alike must recognize that embodied communication technologies are different from text-based services. Their interactions occur in shared physical,  emotional and intimate contexts, involving vulnerability, empathy, and co-presence. Even more so than private text exchanges. Designing digital surveillance frameworks that blindly extend into these domains risks undermining the very benefits that make such technologies valuable: our ability to trust them with our most intimate and vulnerable moments. Rather than folding robots into existing regulatory categories, future governance must explicitly acknowledge their hybrid nature as communicative partners, and social participants. Only then can we ensure that future robots connect rather than control us. 

\section{From Surveillance to Exploitation}

The real-time scanning introduced by Chat Control could go beyond spying for safety, drifting into data theft and behavioural manipulation. As discussed earlier, robots used in care, education, or telepresence may generate not only audio and video, but also detailed behaviour logs and predictive models for personalization or safety checks. This distributed processing and storage architecture enlarges the attack surface by increasing the points where sensitive data is handled, creating more opportunities for attackers to intercept or manipulate information. In this section, we first discuss how surveillance can evolve into data theft and manipulation, and then how such capabilities could escalate into direct control of the robot. 

\subsection{Data Exfiltration Backdoors}

Laws and vendor practices that enforce device scanning or embed hidden access channels effectively create data pipelines that capture and store intimate robot interactions, including audio, video, behavioural logs, and model outputs. These pipelines give attackers or hostile states ready-made entry points, while allowing manufacturers to claim legal compliance. Once open, backdoors can be exploited at multiple levels: at the hardware level (e.g., through hardcoded keys and debug interfaces~\cite{bin4ry_unipwn_2025}) 
and at the reasoning level (e.g., through adversarial triggers in learning or reasoning models, including LLMs~\cite{jiao2024can}). Although some of these access mechanisms have beneficial goals, in practice it is virtually impossible to design a backdoor that admits only legitimate actors; keys can leak, access controls can fail, and learning systems can be tricked.

As discussed earlier, AI-driven robots amplify these risks, especially when attacks are targeted against deep learning models. If attackers obtain access to these datasets or to the models trained on them, they can do more than observe. 
They can turn surveillance into exploitation. For example, attackers can infer private information through model-inversion and membership-inference attacks. Model inversion attacks, first identified in 2015, reveal or reconstruct approximate representations of the data used to train a machine learning model~\cite{song2022survey}. Membership inference attacks determine whether a specific data sample was part of the training set~\cite{hu2022membership}; for example, if an attacker can determine that a particular clinical record was used to train a model associated with a given disease, they can infer with high probability that the record's owner has that disease~\cite{hu2022membership}. In short, surveillance channels designed for safety can become tools for exploitation. Attackers can use legally required data collection to uncover private information, then use it to hack robots, gain control, or manipulate their behaviour.

To mitigate such privacy risks, researchers have explored decentralized learning approaches that avoid central data aggregation. One of the most prominent is federated learning (FL), a machine learning paradigm designed to enable model training across data silos while preserving privacy. First introduced by Google in 2016 to improve text prediction on Android devices~\cite{mcmahan2017communication}, FL allows multiple clients to collaboratively train a shared model under a decentralized framework~\cite{li2020review}. In recent years, the application of FL to cybersecurity for Internet of Things (IoT) has attracted significant research interest~\cite{ghimire2022recent}. However, FL has introduced novel classes of attacks targeting data and model parameters~\cite{ghimire2022recent}. Such attacks can be executed by either forging end devices' local data or manipulating model parameters on the client or server side. Similarly, backdoor-related vulnerabilities may expose robots to attacks in which the compromise of monitoring infrastructure results in real-world harm.

Despite advances such as FL, technical solutions alone cannot guarantee safety or trust. In HRI, these measures must be complemented by ethical safeguards that minimize data collection in vulnerable contexts (e.g., care or therapy) and ensure user-awareness and consent.

\subsection{Control Backdoors}
Recently, researchers have shown that Unitree shipped its commercial humanoids and quadrupeds with hardcoded keys~\cite{bin4ry_unipwn_2025}. The 
manufacturer's hidden access pathways function as built-in backdoors, under the guise of maintenance or user support. 
If regulations akin to Chat Control begin to create incentives for such monitoring capabilities, manufacturers gain the perfect pretext to normalize them. They can justify pervasive data access by invoking legal compliance. This, in turn, allows them to reframe a problematic design choice as a compliance feature, thereby normalizing surveillance as a standard safety mechanism. 

Backdoors discovered in robotic products are not limited to the hardware level; they also span algorithmic layers. Jiao et al.~\cite{jiao2024can} 
demonstrated that robots leveraging large language models (LLMs) are vulnerable to backdoor triggers concealed in words, scenarios, or knowledge fragments, similarly to the underlying models themselves. Such triggers can silently compromise a robot's decision-making and lead to catastrophic behaviours, such as autonomous vehicles accelerating toward obstacles or home robots acting in violation of their intended purpose. As robots increasingly rely on LLMs to interpret human commands and make autonomous decisions~\cite{zhang2023large}, influencing their language provides access to their actions, further blurring the boundary between communication and control. 

This reveals a deeper dimension of the \qt{too safe to be safe} paradox: efforts to make robots more intelligent, adaptive, and socially aware may also make them easier to deceive, coerce, or repurpose. 
Both recent exploits and speculative fiction warn that backdoors that start as convenience or safety mechanisms, can escalate into catastrophic vulnerabilities. Popular culture has already anticipated such scenarios. British TV series Black Mirror's \textit{Hated in the Nation} episode depicted government backdoors in autonomous systems being hijacked to commit mass harm~\cite{brooker2016hated}. This brings us closer to the 
speculative dystopia where robots designed to assist, protect, or entertain become vectors of surveillance, sabotage, or violence if hidden access points are normalized and left unsecured.

If attackers exploit such backdoors, they could take remote control of robots and turn them into spying tools, secretly recording video, audio, or leaking sensitive data. In the worst case, hacked robots could harm vulnerable people, or serve as entry points for deeper hacks into other secure systems. Ross Anderson's~\cite{anderson2022chat} analysis of Chat Control highlights how laws framed as child-protection measures risk normalising mass surveillance by incentivising client-side scanning and weakening encryption. Similarly, Abelson et al.~\cite{abelson2024} argue that client-side scanning neither ensures effective crime prevention nor prevents surveillance. Once embedded into personal devices, such mechanisms effectively turn private technology into extensions of state inspection. 

Current robotic systems have already exhibited various cybersecurity issues~\cite{mayoral2022robot}. If the Chat Control proposal were applied to embodied systems, the boundary between digital surveillance and physical compliance would disappear. Anderson~\cite{anderson2022chat} warns against techno-solutionism, meaning the over-reliance on technology-driven approaches to solve social problems and change social norms~\cite{saetra2024technological}. He emphasises that social and ethical challenges cannot be addressed through automated monitoring but instead require human responsibility. Chat Control places both citizens and machines under observation, establishing an infrastructure of control. 
\section{A Day Under Chat Control}
Once embedded in daily life, detection mechanisms seldom remain confined to their original scope. Although they may begin with the automated detection of illegal content, they can easily drift into identifying so-called \qt{risky} or \qt{harmful} behaviour. Here, we imagine an illustrative HRI scenario in which the same moderation logic that governs online communication platforms extends into embodied contexts.

To illustrate how Chat Control could quietly reshape everyday human–robot relations, consider a near-future care scenario inspired by the film Robot \& Frank~\cite{robotandfrank2012}. What happens when the robots designed to support and comfort us also become obligated to watch and report us? The robot that once comforted its user now sends regular reports of conversations and emotional states to relatives and healthcare personnel. In this world, care becomes inseparable from surveillance, and intimacy is mediated through systems of compliance rather than trust. 

Frank (a lonely older man) is gifted a robot to ease his loneliness. The robot can engage socially, communicate with him, understand his emotions, and respond accordingly. It can also perform household tasks such as cooking and cleaning.  He enjoys the company of his care robot as it helps him at home, reminds him about his medications, and keeps him company. Over time, however, the robot starts to pause more often, hesitating before answering. When Frank jokes about skipping his pills, the robot's light blinks red: \qt{For your safety, this information has been shared with your healthcare provider.} Later, it refuses to open the door, claiming that Frank's request might be unsafe. The robot's calm, kind voice remains the same, but its purpose has changed. What once felt like care now feels like control. Under Chat Control, the robot still protects Frank, but not for him, rather from him. Despite the benefits and the positive impact on his overall well-being, Frank is no longer comfortable having the robot around, knowing that it might report his behaviour.


\section{Conclusion}
In this paper, we have shown how a policy aimed at regulating digital communication could, through legal and technical overreach, extend into the embodied realm of HRI. Through the lens of the Chat Control proposal, the paper demonstrates how a legal framework that treats robots as communication services collapses the boundary between observation and participation. When coupled with advances in AI, such monitoring obligations risk turning robots into cognitive extensions of the state: systems capable of interpreting, classifying, and reporting human behaviour in real time. These architectures invite technical exploitation, moral ambiguity, and long-term erosion of user agency. To prevent this, safety and security must be understood as mutually reinforcing rather than opposing goals. Regulation should prioritise transparency, local-first processing, and robust oversight mechanisms that preserve privacy without undermining protection. Critical HRI must continue to question how laws and technologies shape the ethics of interaction. 

\section{Acknowledgment}
This work has been partially supported by the Wallenberg AI, Autonomous Systems and Software Program (WASP) funded by the Knut and Alice Wallenberg Foundation, and by the Wallenberg AI, Autonomous Systems and Software Program - Humanities and Society (WASPHS) funded by the Marianne and Marcus Wallenberg Foundation and the Marcus and Amalia Wallenberg Foundation.

In preparing this manuscript, the authors used GPT-5 (OpenAI, San Francisco, CA, USA) as an assistive tool for language refinement including improving sentence clarity, enhancing transitions between paragraphs, as well as exploratory ideation. All outputs were critically assessed, substantially revised, and validated by the authors, who retain full responsibility for the conceptual development, methodology, and scientific conclusions of this work.

\bibliography{sn-bibliography}

\end{document}